\documentclass[prl,twocolumn,preprintnumbers,amsmath,amssymb,floatfix,longbibliography]{revtex4}
\usepackage{graphicx}
\usepackage{graphics}
\usepackage{psfrag}
\usepackage{hyperref}
\usepackage{multirow}
\usepackage[sort&compress]{natbib}
\usepackage{amssymb}
\usepackage{amsmath}
\usepackage{amsthm}
\usepackage{bbold}
\usepackage{bbm}
\usepackage{enumerate}

\newcommand{\vphi}{\varphi}

\newcommand{\vare}{\varepsilon}
\newcommand{\sgn}{\mbox{sgn}}
\newcommand{\rmi}{{\rm i}}
\newcommand{\mathJ}{\mathcal{J}}

\usepackage{xcolor}

\begin{document}

\hypersetup{pdftitle={title}}
\title{Interplay of topology and electron-electron interactions in\\ Rarita--Schwinger--Weyl semimetals}

\author{Igor Boettcher}
\email{iboettch@umd.edu}
\affiliation{Joint Quantum Institute, University of Maryland, College Park, MD 20742, USA}

\begin{abstract}
We study for the first time the effects of strong short-range electron-electron interactions in generic Rarita--Schwinger--Weyl semimetals hosting spin-3/2 electrons with linear dispersion at a four-fold band crossing point. The emergence of this novel quasiparticle, which is absent in high-energy physics, has recently been confirmed experimentally in the solid state. We combine symmetry considerations and a perturbative renormalization group analysis to discern three interacting phases that are prone to emerge in the strongly correlated regime: The chiral topological semimetal breaks a $\mathbb{Z}_2$-symmetry and features four Weyl nodes of monopole charge +1 located at vertices of a tetrahedron in momentum space. The s-wave superconducting state opens a Majorana mass gap for the fermions and is the leading superconducting instability. The Weyl semimetal phase removes the fourfold degeneracy and creates two Weyl nodes with either equal or opposite chirality depending on the anisotropy of the band structure. We find that symmetry breaking occurs at weaker coupling if the total monopole charge remains constant across the transition.
\end{abstract}

\maketitle

The emergence of massless fermionic quasiparticles as low-energy degrees of freedom in condensed matter systems links phenomena from high-energy physics to those of many-body systems \cite{BookVolovik}. Semimetals with the Fermi level close to a high-symmetry band crossing point provide the closest realization of the relativistic concept of a particle described by its mass and spin \cite{BookWeinberg}. The exploration of such Fermi points in graphene, ultracold atoms, Dirac, Weyl, and Luttinger semimetals is on the forefront of both theoretical and experimental research \cite{GeimReview,RevModPhys.82.3045,RevModPhys.83.1057,NatureDiracSOC,PhysRevLett.119.206401,PhysRevLett.119.206402,RevModPhys.90.015001,PhysRevB.97.241101,PhysRevLett.121.157602,GaoReview}.

\begin{figure}[t]
\centering
\includegraphics[width=4.51cm]{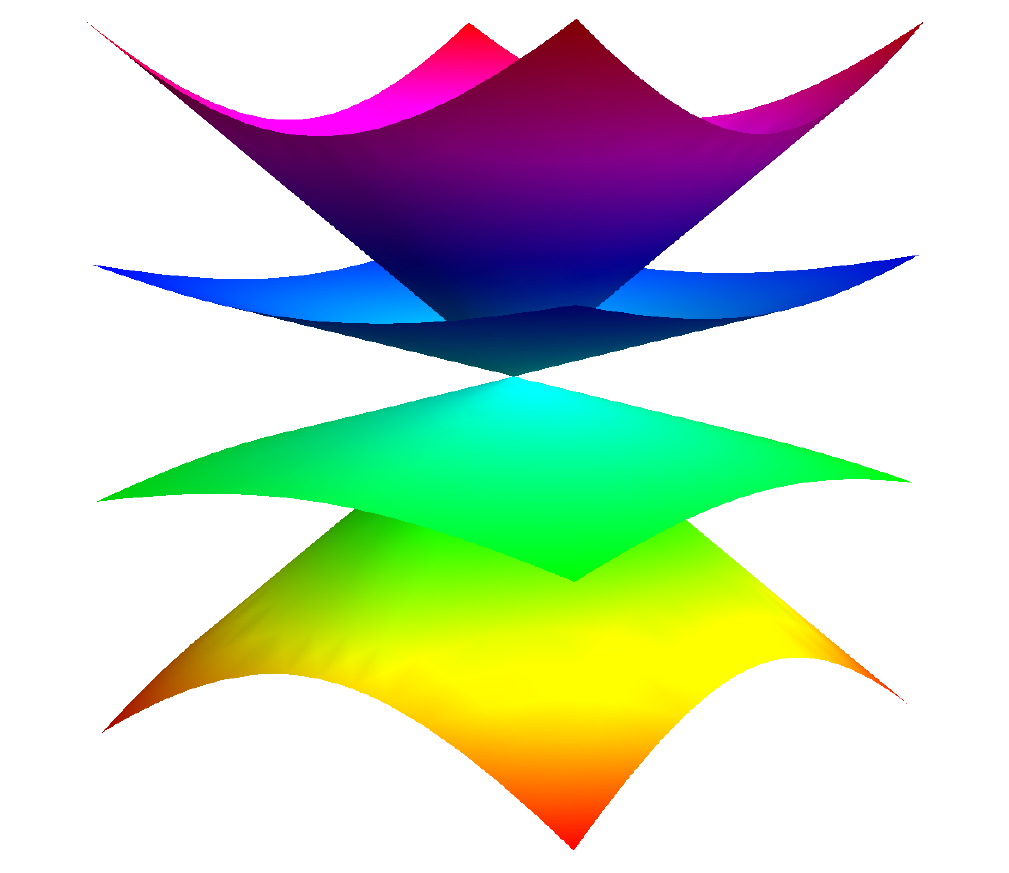} 
\includegraphics[width=3.9cm]{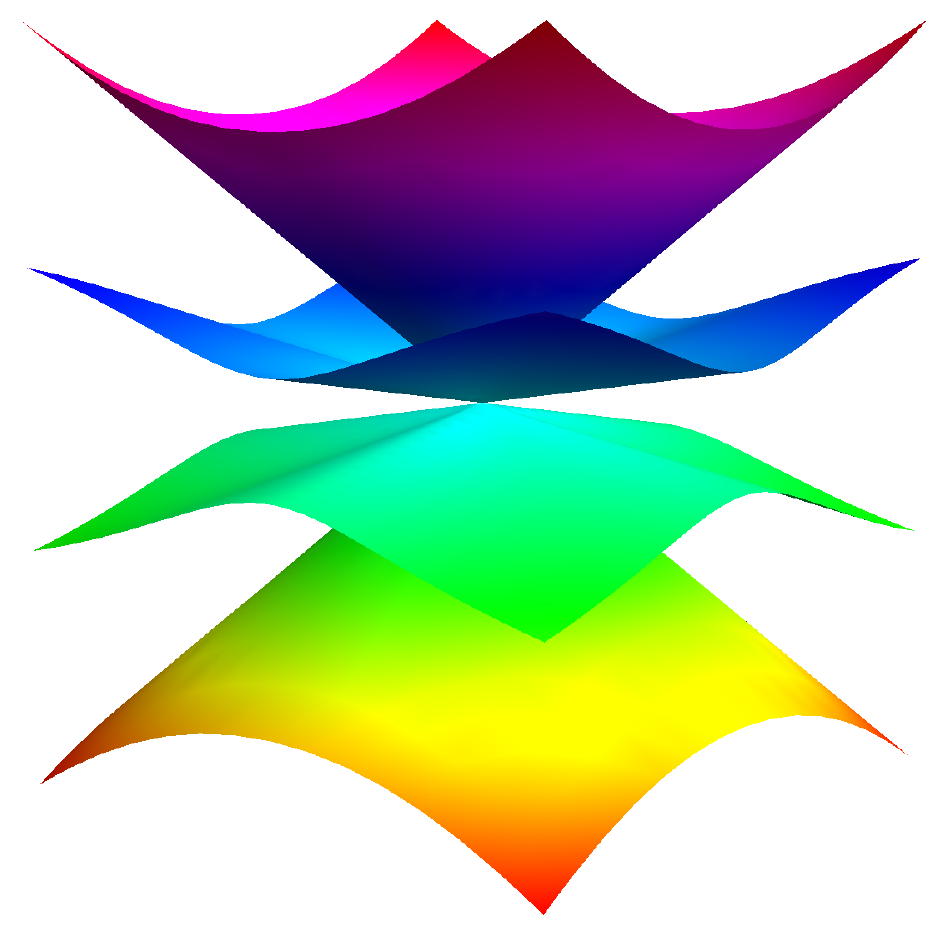}
\caption{Energy dispersion of the spin-3/2 Rarita--Schwinger--Weyl fermion at a four-fold linear band crossing point. We set one of the momentum components of $\textbf{p}$ to zero and plot the eigenvalues of Eq. (\ref{intro1}) versus the remaining two components for $v_2=0$ (left) and $v_2=1/3$ (right). For $v_2=0$ the spectrum is rotation invariant, while for generic values of $v_2$ it features cubic anisotropy.}
\label{FigDisp}
\end{figure}

Very recently, first experimental evidences of emergent spin-3/2 relativistic fermions with concomitantly large topological charge have been reported in CoSi, RhSi \cite{PhysRevLett.122.076402,Rao2019ObservationOU,Sanchez}, AlPt \cite{SchroeterNature}, and PdBiSe \cite{PhysRevB.99.241104}. Since the standard model of particles does not feature fundamental spin-3/2 particles, although they appear as composite degrees of freedom through $\Delta$-baryons or in conjectured extensions like supergravity \cite{PhysRev.60.61,PhysRevD.13.3214,PhysRevB.93.045113}, identifying their condensed matter analogues is key to studying their properties and interactions. In three-dimensional Rarita--Schwinger--Weyl (RSW) semimetals with fourfold linear band crossing point at the Fermi level, the universal low-energy $k\cdot p$ Hamiltonian reads
\begin{align}
 \label{intro1} H(\textbf{p}) = p_i(v_1 J_i + v_2 J_i^3).
\end{align}
Here $\textbf{p}$ is the momentum measured from the crossing point, $J_i$ are the $4\times 4$ spin-3/2 matrices \cite{som}, $i=1,2,3=x,y,z$ with implicit summation over repeated indices, and $v_{1,2}$ are two non-universal material parameters. The term multiplying $v_1$ is rotationally invariant and proportional to the helicity operator with eigenvalues $\pm 3/2, \pm 1/2$, making the spin-3/2 character explicit. The second term is the other scalar (linear in $p_i$) that can be constructed from the cubic group and reduces rotational symmetry to the rotational cubic group $O$ for $v_2\neq 0$, see Fig. \ref{FigDisp}. Concrete candidate materials for realizing $H(\textbf{p})$ have been proposed at the transition to a crystalline topological insulator in antiperovskites \cite{PhysRevB.90.081112,PhysRevB.93.241113}, for many space groups and materials in Refs. \cite{Bradlynaaf5037,chang2018topological,2019arXiv190412867C}, in transition metal silicides \cite{PhysRevLett.119.206402}, and for $v_2=0$ through a specific tight-binding model with isotropic spin-orbit coupling on a tricolor lattice in Ref. \cite{PhysRevB.94.195205}. 
Our model in Eq. (\ref{intro1}) is idealized in the sense that we do not assume other band crossings at the Fermi level to be important for the interacting phases, including intervalley coupling to an RSW fermion of opposite chirality.

The impact of short-range interactions in generic RSW semimetals has not been studied so far. This is somewhat surprising when compared to the case of \emph{quadratic} band touching of spin-3/2 electrons, with Eq. (\ref{intro1}) replaced by the Luttinger Hamiltonian \cite{moon,PhysRevLett.113.106401}, where material realizations in pyrochlore iridates and half-Heuslers are rather well-understood, and there exists an extensive literature on exotic interacting phases resulting from the higher spin of fermions such as spin-2 or spin-3 Cooper pairing \cite{PhysRevB.93.205138,PhysRevLett.116.177001,PhysRevLett.118.127001,PhysRevB.95.144503,PhysRevB.96.144514,PhysRevB.96.214514,PhysRevLett.120.057002,PhysRevX.8.011029,Kimeaao4513,PhysRevB.97.205402,Mandal,PhysRevB.98.104514,PhysRevB.99.054505,2018arXiv181104046S} or octupolar magnetism \cite{PhysRevX.4.041027,WKrempa,PhysRevB.92.035137,PhysRevB.95.085120,PhysRevB.95.075149,PhysRevX.8.041039}. In both RSW and Luttinger semimetals, weak short range interactions are irrelevant due to the vanishing density of states at the Fermi point, and so the phases of interest are at strong coupling. For RSW semimetals, short-range interactions have only been investigated in the exceptional case of $\alpha=0$ \cite{PhysRevB.97.241101,PhysRevLett.121.157602}(defined below), which is qualitatively different from $H=p_iJ_i$.

In this Letter we aim to fill this gap in the understanding of interacting RSW semimetals. Our analysis proceeds in three steps. We first study the single-particle physics of RSW fermions to clarify the distinct relevant parameter regimes. We then perform an unbiased perturbative renormalization group (RG) analysis of all competing ordering channels and identify three leading strong coupling instabilities. At last we discuss symmetries and quasiparticle spectra in the ordered phases found with the RG.

\emph{Single-particle physics.} To discuss the symmetries and topology of the RSW Hamiltonian in Eq. (\ref{intro1}), we write
\begin{align}
 \label{mod4} H(\textbf{p}) = p_i(V_i+\alpha U_i)
\end{align}
with $V_i = \frac{1}{3}(-7J_i+4J_i^3)$ and $ U_i =\frac{1}{6}(13 J_i-4J_i^3)$ \cite{PhysRevB.93.241113}. We have $\mbox{tr}(V_iV_j)=\mbox{tr}(U_iU_j)=4\delta_{ij}$ and $\mbox{tr}(V_iU_j)=0$. The chemical potential is at the band crossing point. We set the Fermi velocity multiplying the term $p_iV_i$ to unity so that the crossing is described by the single parameter $\alpha$ \footnote{In the notation of Eq. (S121) of the supplemental material of Ref. \cite{Bradlynaaf5037} we have $\alpha=(a-b)/(a+b)$.}. For $\alpha=2$ the Hamiltonian reduces to the rotationally invariant expression $p_iJ_i$. Remarkably, the matrices $V_i$ form a Clifford algebra,
\begin{align}
 \label{band5b} \{V_i,V_j\}= 2\delta_{ij},
\end{align}
and so $H_{\alpha=0}=p_iV_i$ is Lorentz invariant with enhanced $\text{O}(2)$-symmetry. Importantly, this comprises two Weyl points of \emph{equal} chirality, contrary to a Dirac Hamiltonian, which decomposes into Weyl points of opposite chirality in the massless limit.

The system is time-reversal invariant with time-reversal operator $\mathcal{T} = \gamma_{45}\mathcal{K}$, $\mathcal{T}^2=-1$, where $\mathcal{K}$ is complex conjugation and $\gamma_{45}$ a Hermitean matrix defined below. For fixed $\textbf{p}$ we have $\{\mathcal{T},H(\textbf{p})\}=0$, and so every eigenvalue $E(\textbf{p})$ implies an eigenvalue $-E(\textbf{p})$ for the time-reversed eigenvector, i.e. particle-hole symmetry of the spectrum. Next consider the Hermitean operator
\begin{align}
 \label{mod11} \mathcal{W} = \frac{2}{\sqrt{3}}(J_xJ_yJ_z+J_zJ_yJ_x),
\end{align}
which squares to unity. We have $[V_i,\mathcal{W}]= \{U_i,\mathcal{W}\}=0$, implying $\mathcal{W} H_\alpha\mathcal{W} = H_{-\alpha}$. Consequently a sign change $\alpha\to -\alpha$ can be undone by $\psi \to \mathcal{W}\psi$, and so we assume $\alpha\geq 0$.

\renewcommand{\arraystretch}{1.6}
\begin{table}[t]
\begin{tabular}{|c|c|c|c|c|}
\hline \ band \ & energy & $\alpha=0$ & \ $0 < \alpha <1$ \ & $1<\alpha$ \\
\hline $1$ & $E_+(\textbf{p})$ & \ $C=-1$ \ &  $C=3$  &  $C=3$  \\
 \hline $2$ & $E_-(\textbf{p})$ & $C=-1$ & $C=-5$ & $C=1$ \\
\hline $3$ & \ $-E_-(\textbf{p})$ \ & $C=1$ & \ $C=5$ \ & \ $C=-1$ \ \\
\hline $4$ & $-E_+(\textbf{p})$ & $C=1$ & $C=-3$ & $C=-3$\\
 \hline \multicolumn{2}{|c|}{\ monopole charge \ } & $-2$ & $-2$ & $4$ \\
\hline
\end{tabular}
\caption{Normal state Chern numbers. Bands are enumerated by decreasing energy eigenvalues, see Eq. (\ref{mf5}) with $\chi=0$. There is a topological phase transition at $\alpha=1$, where the total monopole charge changes. The case $\alpha=0$ corresponds to two overlapping Weyl nodes of equal chirality.}
\label{TabChern}
\end{table}
\renewcommand{\arraystretch}{1}

We now determine the topology of the RSW point node. The eigenvectors $|\nu(\textbf{p})\rangle$ of $H(\textbf{p})$ for fixed $\textbf{p}$ comprise two positive and two negative energy bands, which we label by an index $\nu$. For each band we define the Berry connection $\textbf{A}_\nu(\textbf{p}) = -\rmi \langle \nu(\textbf{p})| \nabla_{\textbf{p}}|\nu(\textbf{p})\rangle$, pseudo-magnetic field $\textbf{B}_\nu(\textbf{p}) = \nabla_{\textbf{p}} \times \textbf{A}_\nu(\textbf{p})$, and Chern number $C_\nu=\oint \mbox{d}\vec{\Omega}\cdot \textbf{B}_\nu(\textbf{p})$, where the latter surface integral encloses the origin. In Table \ref{TabChern} we present $C_\nu$ as a function of $\alpha$ together with the total monopole charge of the Fermi node, defined as the sum of Chern numbers of the positive energy bands. The system undergoes a topological phase transition at $\alpha=1$, where the monopole charge changes from $-2$ to $4$. (Note that $H(\textbf{p})$ features line nodes for $\alpha=1$, which are an artefact of the linear approximation and can be eliminated by including a quadratic term.) 
The curious Chern numbers in the regime $0<\alpha<1$ do not seem to have been reported before. Note that while RSW fermions are often associated with monopole charge 4 as in AlPt or PdBiSe, the case of charge 2 observed in CoSi/RhSi might also correspond to an RSW fermions.

\emph{Renormalization group.} The many-body physics of interacting RSW electrons is captured by the Lagrangian
\begin{align}
 \label{mod6} L = \psi^\dagger(\partial_\tau +H(-\rmi \nabla)+\bar{e}a)\psi +\frac{1}{2}(\nabla a)^2 + L_{\rm short}
\end{align}
with $\tau$ imaginary time, $\psi$ the four-component electron spinor, $a$ the photon that mediates long-range interactions, $\bar{e}$ electric charge, and short-range interactions $L_{\rm short}$. Due to the symmetries and dimensionality of the system, the interaction effects at low energies can be investigated with the one-loop RG developed in Ref.  \cite{PhysRevB.95.075149}, where Eqs. (\ref{mod6}) and (\ref{mod7}) have been investigated for $H$ being the Luttinger Hamiltonian. Here we use the same notation and adapted equations, but obviously the physics is different due to the modified band dispersion. We refer to Ref.  \cite{PhysRevB.95.075149} for a very detailed introduction to the computational procedure, but summarize a few central definitions in the Supplemental Material (SM, \cite{som}).

In the following, we only need to consider point-like short-range interactions, since terms containing derivatives of the fermion field are suppressed at the Fermi point at low energies. The most general Fierz-complete form is \cite{PhysRevB.95.075149}
\begin{align}
\label{mod7} L_{\rm short} = \bar{g}_1 (\psi^\dagger\psi)^2 + \bar{g}_2 \sum_{a=1}^2(\psi^\dagger \gamma_a \psi)^2 + \bar{g}_3\sum_{a=3}^5(\psi^\dagger \gamma_a\psi)^2,
\end{align}
where we introduce five $\gamma$-matrices
\begin{align}
  \label{mod8}  \gamma_1 &= \frac{J_x^2-J_y^2}{\sqrt{3}},\ \gamma_2= J_z^2-\frac{5}{4}\mathbb{1},\ \gamma_3=\frac{J_zJ_x+J_xJ_z}{\sqrt{3}},\\
  \label{mod9} \gamma_4 &= \frac{J_yJ_z+J_zJ_y}{\sqrt{3}},\ \gamma_5=\frac{J_xJ_y+J_yJ_x}{\sqrt{3}}
\end{align}
satisfying $\{\gamma_a,\gamma_b\}=2\delta_{ab}$. We write $\gamma_{ab}=\rmi \gamma_a\gamma_b$. For the RG analysis, we define dimensionless running couplings by $g_i = \Lambda^2\bar{g}_i/(2\pi^2) ,\ e^2 =\bar{e}^2/(2\pi^2)$, with $\Lambda$ the bandwidth. Although the $g_i$ are power-counting irrelevant, they can induce ordering at strong coupling \cite{herbutbook}.

As pointed out by Isobe and Fu \cite{PhysRevB.93.241113}, the electric charge $e$ gives self-energy corrections, but flows to zero. It leads to an anomalous dimension $\propto e^2$ for the fermions and, remarkably, the stable fixed points for the anisotropy are $\alpha=0$ and $\alpha=2.296$, whereas $\alpha=2$ is unstable. In real materials, however, the corresponding RG flow may be stopped by finite volume effects, or suppressed by a large dielectric constant. We thus assume $\alpha$ to be a fixed number, determined by the chemical composition of the compound. Renormalization effects due to the coupling of long- and short-range interactions are equally suppressed by powers of $e^2\to 0$ and will be neglected henceforth. (Furthermore, there is no one-loop diagram that could induce a back-reaction of $g_{1,2,3}$ onto the RG flow of $e^2$ \cite{PhysRevB.95.075149}.)
The remaining RG flow equations have the form $\mbox{d} g_i/\mbox{d}\ln b = -2g_i + C_{ijk}(\alpha) g_j g_k$, where $C_{ijk}(\alpha)$ are coefficients that result from integrating out fluctuations of RSW electrons in a momentum shell $\Lambda\geq p \geq \Lambda/b$. The coefficients parametrically depend on $\alpha$ through the anisotropic fermion dispersion.

We search for quantum critical points, which are fixed points of the RG flow where exactly one linear combination of $g_1,g_2,g_3$ is a relevant direction. At every fixed point, we determine the scaling dimension of the ten fermion bilinears $\psi^\dagger M \psi^{(*)}$ allowed by symmetry through coupling a term $h \psi^\dagger M \psi^{(*)}$ to the Lagrangian and determining the flow $\dot{h} =(1+\eta) h$. The bilinear with the largest susceptibility $\eta$ condenses at the associated quantum phase transition \cite{som}. Both the fixed points and susceptibilities depend on $\alpha$. We identify three distinct quantum critical points (labelled W, SC, V), which are related to the following order parameters:
\begin{itemize}
 \item[(1)] chiral topological semimetal: $\chi=\langle\psi^\dagger \mathcal{W}\psi\rangle\neq 0$
 \item[(2)] s-wave superconductor: $\phi=\langle \psi^\dagger \gamma_{45}\psi^*\rangle\neq0$
 \item[(3)] Weyl semimetal: $m_i=\langle \psi^\dagger V_i\psi\rangle\neq 0$
\end{itemize}
The identification of these three leading instabilities in interacting RSW semimetals from an unbiased RG analysis constitutes the first major result of this work.

\begin{figure}[t!]
\centering
\begin{minipage}{0.48\textwidth}
\includegraphics[width=7cm]{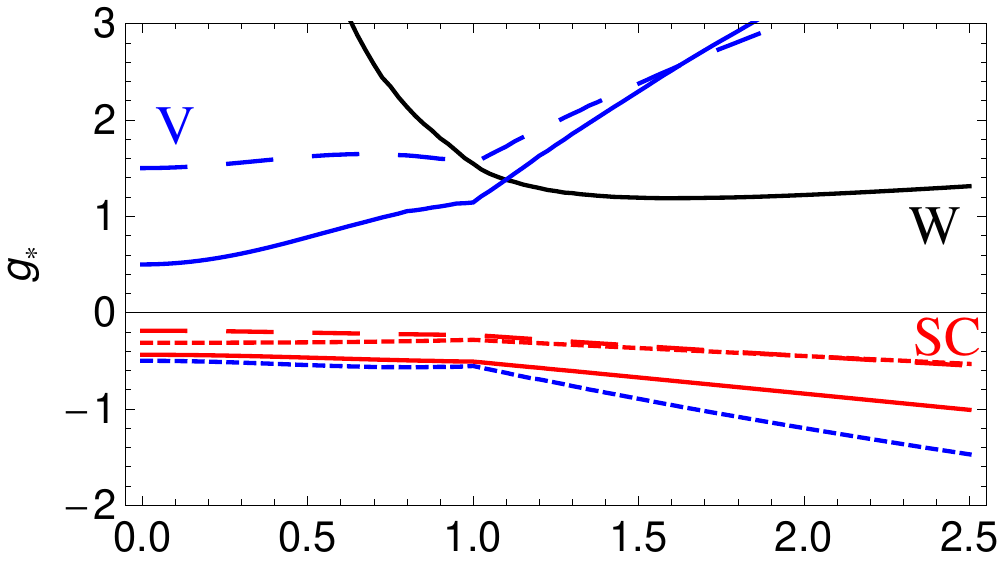}
\includegraphics[width=7cm]{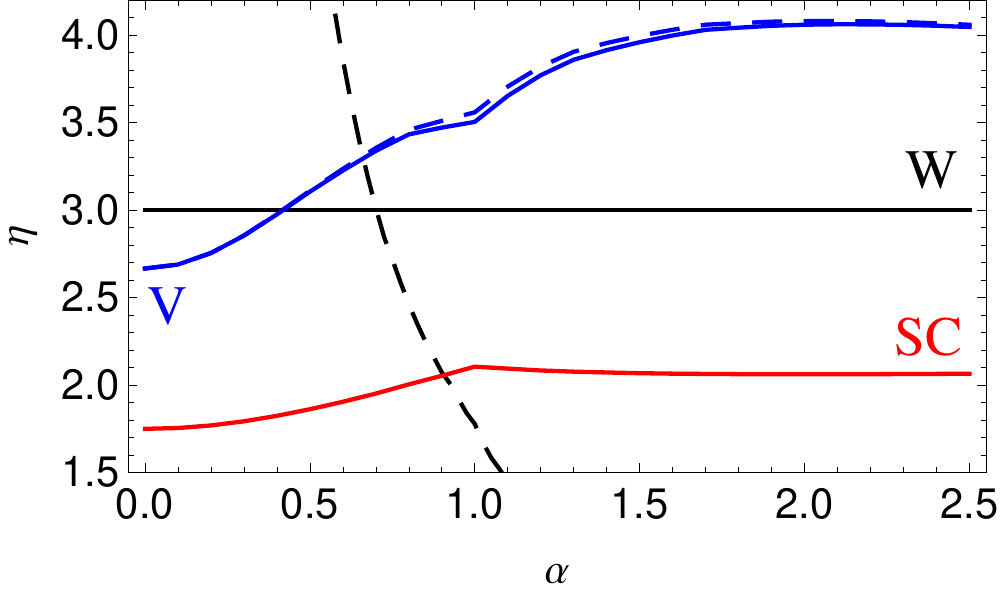}
\caption{Renormalization group fixed points. \emph{Upper panel.} Couplings $g_1,g_2,g_3$ (solid, dashed, dotted line) at the fixed points W (black), SC (red), and V (blue). At W we have $g_1=-g_2=g_3$ and so we only show the first coupling. \emph{Lower panel.} Susceptibility exponent $\eta$ of the order parameter at the fixed points. At W we have $\eta=3$ for $\langle \psi^\dagger \mathcal{W}\psi\rangle$ for all $\alpha>0$ (solid black). This is the dominant divergence for $\alpha>0.7$, while for $\alpha<0.7$ the order parameter $N_i(\alpha)=\langle\psi^\dagger(V_i+\kappa(\alpha) U_i)\psi\rangle$ with $\kappa(\alpha)\gg1$ has the largest susceptibility (dashed black). At SC the most divergent channel is the s-wave superconductor (red). At V, $N_i(\alpha)$ has the largest susceptibility exponent (dashed blue). Since here $\kappa(\alpha)$ is generically very small, we can neglect it and only consider the approximate order parameter $\langle\psi^\dagger V_i\psi\rangle$ (solid blue).}
\label{FigRG}
\end{minipage}
\end{figure}

The three fixed points have the following properties, which are visualized in Fig. \ref{FigRG}: The large critical couplings $g_{i,\rm c} \sim 1$ are due to the vanishing density of states at the Fermi point. Whereas SC and V exist for every $\alpha$, W only exists for $\alpha>0$. \underline{W}: Here the ratio of critical couplings is $g_1=-g_2=g_3=2g_\star>0$ for all $\alpha$, implying the system to flow to the fixed point Lagrangian (\ref{mf1}). The susceptibility exponent of $\chi$ is exactly given by the spatial dimension, $\eta_{\mathcal{W}}=d=3$, which comprises the leading instability in the regime $\alpha\geq 0.70$. For $\alpha<0.7$, the order parameter instead has large overlap with $\langle \psi^\dagger U_i \psi\rangle$, but we will not further discuss it in this work. \underline{SC}: This is a superconducting quantum critical point corresponding to a condensation of $\phi$, which acts as a Majorana mass term for the fermions, see Eq. (\ref{mass5}). \underline{V}: The fixed point V corresponds to a condensation of $m_i$. More precisely, the order parameter receives a small admixture of $U_i$ according to $\langle \psi^\dagger (V_i+\kappa(\alpha) U_i)\psi\rangle$. However, $\kappa=0$ for $\alpha=0$ and $\kappa(\alpha) < 5\%$ in general, so we neglect this effect for the discussion here, but plot the function $\kappa(\alpha)$ for completeness in the SM \cite{som}.

\emph{Chiral topological semimetal.} We now characterize the properties of the chiral topological semimetal phase, which is the second major result of this work. We verified above that the system at W, after fine-tuning one coupling, is attracted to the fixed point Lagrangian
\begin{align}
 \label{mf1} L_{\star} = \psi^\dagger(\partial_\tau +H)\psi - \bar{g}_\star (\psi^\dagger\mathcal{W}\psi)^2
\end{align}
with $\bar{g}_\star>0$. In the mean field approximation we replace $-\bar{g}_\star(\psi^\dagger\mathcal{W}\psi)^2 \to \chi (\psi^\dagger \mathcal{W}\psi)$ and arrive at effectively noninteracting fermionic quasiparticles described by the Hamiltonian $H_{\rm mf}(\textbf{p})=H(\textbf{p}) +\chi\mathcal{W}$. Note that  $\mathcal{W}$ is invariant under the rotational or \emph{chiral} tetrahedral group $T$ only \cite{som}. The term $\chi \mathcal{W}$ breaks time-reversal symmetry, but preserves particle-hole symmetry of the energy spectrum. The positive eigenvalues of $H_{\rm mf}(\textbf{p})$ are
\begin{align}
 \nonumber &E_{\pm}(\textbf{p}) = \Biggl[\chi^2+(1+\alpha^2)p^2\pm\Biggl(4\chi^2p^2+\alpha^2\Bigl[4p^4\\
 \label{mf5} &{}-3(4-\alpha^2)\sum_{i<j}p_i^2p_j^2+12\sqrt{3}\chi p_1p_2p_3\Bigr]\Biggr)^{1/2}\Biggr]^{1/2}.
\end{align}
We have $E_+(\textbf{p})>0$. The zeros of $E_-(\textbf{p})$ are located at the four vertices of a tetrahedron according to $\textbf{p}_n=(\chi/\sqrt{3})\textbf{e}_n$ with
\begin{align}
  \label{mf6} \textbf{e}_1&=  \begin{pmatrix} 1 \\ 1 \\ 1 \end{pmatrix},\ \textbf{e}_2= \begin{pmatrix} -1 \\ -1 \\ 1 \end{pmatrix},\  \textbf{e}_3 =\begin{pmatrix} -1 \\ 1 \\ -1 \end{pmatrix},\ \textbf{e}_4= \begin{pmatrix} 1 \\ -1 \\ -1 \end{pmatrix}.
\end{align}
The $\alpha$-dependence of the nodes is implicit, through $\chi$, which is the solution of an $\alpha$-dependent gap equation. The sign of the order parameter $\chi$ gives the configuration of Weyl nodes a handedness so that it cannot be rotated into its mirror image with z-component reversed in sign, thereby breaking a $\mathbb{Z}_2$ symmetry.


To clarify the nature of the gapless quasiparticles at the nodal points, we compute $\textbf{B}_\nu(\textbf{p})$ for the two bands with energy $\pm E_-(\textbf{p})$ and determine the Chern number $C$ from the surface integral surrounding $\textbf{p}_n$ in momentum space. At each vertex of the tetrahedron the positive energy band has $C=1$ and the negative energy band has $C=-1$, so the total monopole charge is $4$. Thus for $\alpha>1$ the phase transition is such that the normal state charge of $+4$ is distributed onto four unit charges $+1$. For $0<\alpha<1$, on the other hand, symmetry breaking implies a topological phase transition which changes the total monopole charge.

The effective Weyl Hamiltonian that describes excitations with momentum $\textbf{p}=\textbf{p}_n+\delta \textbf{p}$ close to the nodal points can be obtained from projecting onto the subspace spanned by the zero modes $|0_n\rangle, |0_n'\rangle$ of $H(\textbf{p}_n)$, yielding $H_0^{(n)}(\delta\textbf{p}) = v^{(n)}_{ij}\delta p_i \sigma_j$, which constitutes type-I Weyl nodes. The energy close to the nodal point reads $E^{(n)}(\delta \textbf{p}) = \pm \sqrt{\delta p_i (v^{(n)}v^{(n)T})_{ij}\delta p_j}$ and the monopole charge of each Weyl node is consistently given by $\sgn[\mbox{det}(v^{(n)})] = \sgn(\alpha^2)=1$. The matrices $v^{(n)}$ are displayed in the supplemental material (SM) \cite{som}.

\emph{Dirac, Majorana, and Weyl mass terms.} Identifying fermion bilinears that open a full gap (''mass terms'') is an important step in finding energetically favorable ordering patterns for any new single-particle Hamiltonian and as such complements the perturbative RG analysis. In the following we discuss three canonical mass terms in the systems: those of Dirac-, Majorana-, and Weyl-type. We first derive the negative result that the Hamiltonian $H=p_iV_i$ for $\alpha=0$ does not permit a Dirac mass term, which would be a fourth matrix $M$ that anticommutes with all $V_i$. Indeed, the $4\times 4$ Clifford algebra $\{A_n,A_m\}=2\delta_{nm}\mathbb{1}$ allows for two inequivalent representations: One reads $A_i = \mathbb{1}_2\otimes \sigma_i$, which is a reducible representation where no fourth anti-commuting matrix exists. The second solution is $ A_1 = \sigma_1\otimes \mathbb{1}_2,\ A_2=\sigma_3\otimes \mathbb{1}_2,\ A_3=\sigma_2\otimes\sigma_2,\ A_4=\sigma_2 \otimes \sigma_1,\ A_5=\sigma_2 \otimes \sigma_3$, and so after choosing three matrices to construct a Hamiltonian $p_iA_i$, there are two left to form mass terms. With a suitable basis change one easily sees that $V_i \sim \mathbb{1}_2 \otimes \sigma_i$ in RSW semimetals \cite{som}, which is of the first type, implying the leading (particle-number conserving) instability for $\alpha=0$ to have nodes. Note that the Hamiltonian considered in Ref. \cite{PhysRevLett.121.157602} reads $p_i(\mathbb{1}_2\otimes\sigma_i+\beta \sigma_i\otimes \mathbb{1}_2)$, with $\beta$ a real parameter, and so only for $\beta=0$ has overlap with the RSW Hamiltonian.

In the s-wave superconducting phase, the system develops a Majorana mass term. The corresponding effective Lagrangian reads \cite{PhysRevB.93.205138}
\begin{align}
 \label{mass5} L_{\rm sc} =  \psi^\dagger (\partial_\tau +H)\psi - g_{\rm s} (\psi^\dagger \gamma_{45}\psi^*)(\psi^T \gamma_{45}\psi)
\end{align}
with superconducting gap $\Delta \propto \langle\psi^\dagger\gamma_{45}\psi^*\rangle$ and $g_{\rm s}>0$. Recall that a Majorana mass term for two-component fermions reads $\psi^\dagger \sigma_2\psi^*$. The energies of quasiparticles are $E(\textbf{p}) = \pm \sqrt{E_0(\textbf{p})^2+|\Delta|^2}$, with $E_0(\textbf{p})$ the spectrum of $H$. The suppression of low-energy excitations explains the superiority of the s-wave superconductor among all particle-number non-conserving orders in the perturbative RG computation.

Eventually consider adding a Weyl mass $m_iV_i$ to the Hamiltonian. As is well-known, for $\alpha=0$ this merely shifts the position of the Weyl nodes. But for $\alpha>0$, the effect is far more intriguing. Assume the minimal free energy is obtained for a state with residual $\text{SO}(2)$-symmetry and so $\vec{m}=(0,0,m)$. The nodes of the mean-field Hamiltonian $H_V(\textbf{p})=H(\textbf{p})+mV_3$ are located at
\begin{align}
 \label{mass6} \textbf{p}_{\rm a}= \frac{-m}{1+\alpha}(0,\ 0,\ 1)^T,\ \textbf{p}_{\rm b} = \frac{-m}{1-\alpha}(0,\ 0,\ 1)^T,
\end{align}
assuming $\alpha \neq 1$. These momenta again correspond to type-I Weyl nodes \cite{som}. Remarkably, the monopole charge associated to each of the two Weyl nodes is given by
\begin{align}
 \label{mass 7} q_{\rm a} &= -1,\ q_{\rm b} = \sgn(\alpha-1).
\end{align}
Consequently, there is a topological phase transition in the broken phase when crossing $\alpha=1$, with the total monopole charge being $-2$ for $\alpha<1$ and $0$ for $\alpha>1$. For $\alpha<1$ the monopole charge remains constant upon condensation of $m\neq 0$. The identification of this Weyl semimetal phase constitutes the third major result of this work.

\emph{Conclusion.} Our analysis reveals an intriguing interplay between topology and interactions. First, the critical couplings of W and V are smaller in those regimes where the total monopole charge does not change across the transition ($\alpha>1$ for W and $\alpha<1$ for V), and so no topological phase transition occurs besides the symmetry breaking. Second, the critical coupling for W has no kink at $\alpha=1$ and the scaling dimension of the order parameter is independent at $\alpha$, indicating a topological nature of the ordering. The rearranged monopole structure in the ordered phases can be revealed experimentally through surface state spectroscopy \cite{PhysRevLett.119.206401,PhysRevLett.119.206402} or optical response measurements \cite{PhysRevB.99.125146,PhysRevLett.121.157602}. It will be exciting to study the interplay of a pair of RSW fermions with opposite monopole charge, similar to the interplay of Weyl nodes in Weyl semimetals \cite{PhysRevB.90.035126}.

\acknowledgements \emph{Acknowledgements.} I gratefully acknowledge collaboration with Michael Scherer in an early stage of this work. I thank Igor Herbut for inspiring discussions and for bringing to my attention the relation of these findings to the representation theory of Clifford algebras. I thank Fabian von Rohr for insightful comments. This work was supported by DoE BES Materials and Chemical Sciences Research for Quantum Information Science program, NSF Ideas Lab on Quantum Computing, DoE ASCR Quantum Testbed Pathfinder program, ARO MURI, ARL CDQI, and NSF PFC at JQI.

\bibliographystyle{apsrev4-1}
\bibliography{refs_opt}

\cleardoublepage

\setcounter{equation}{0}
\renewcommand{\theequation}{S\arabic{equation}}

\begin{center}
\textbf{\Large Supplemental Material}
\end{center}

\section{Spin-3/2 matrices}\label{AppMat}
The spin-3/2 matrices in their standard matrix representation read
\begin{align}
 \label{mat1} J_x &= \begin{pmatrix} 0 & \frac{\sqrt{3}}{2} & 0 & 0 \\ \frac{\sqrt{3}}{2} & 0 & 1 & 0 \\ 0 & 1 & 0 & \frac{\sqrt{3}}{2} \\ 0 & 0 & \frac{\sqrt{3}}{2} & 0 \end{pmatrix},\\
 \label{mat2} J_y &=\begin{pmatrix} 0 & -\rmi \frac{\sqrt{3}}{2} & 0 & 0 \\ \rmi \frac{\sqrt{3}}{2} & 0 & -\rmi &0 \\ 0 & \rmi & 0 & -\rmi \frac{\sqrt{3}}{2} \\ 0 & 0 & \rmi \frac{\sqrt{3}}{2} & 0\end{pmatrix},\\
 \label{mat3} J_z &= \begin{pmatrix} \frac{3}{2} & 0 & 0 & 0 \\ 0 & \frac{1}{2} & 0 & 0 \\ 0 & 0 & -\frac{1}{2} & 0 \\ 0 & 0 & 0 & -\frac{3}{2}\end{pmatrix}.
\end{align}
The matrices satisfy $[J_i,J_j]=\rmi \vare_{ijk}J_k$ and $\sum_i J_i^2 = \frac{15}{4} \mathbb{1}_4$, and all results obtained in the main text result from these relations. Some insight into the operators that appear in the analysis can be gained from applying the basis change
\begin{align}
 \label{mat4} S = \begin{pmatrix} 1 & 0 & 0 & 0 \\ 0 & 0 & 0 & 1 \\ 0 & 0 & 1 & 0 \\ 0 & 1 & 0 & 0 \end{pmatrix}
\end{align}
with $S^{-1}=S ^T=S$. The spin-3/2 matrices in this frame (denoted with an overbar) read
\begin{align}
 \label{mat5} \bar{J}_x &= SJ_x S = \begin{pmatrix} 0 & 0 & 0 &\frac{\sqrt{3}}{2} \\ 0 & 0 & \frac{\sqrt{3}}{2} & 0 \\ 0 & \frac{\sqrt{3}}{2} & 0 & 1 \\ \frac{\sqrt{3}}{2} & 0 & 1 & 0 \end{pmatrix},\\
 \label{mat6} \bar{J}_y &= SJ_y S = \begin{pmatrix} 0 & 0 & 0 &-\frac{\sqrt{3}}{2}\rmi \\ 0 & 0 & \frac{\sqrt{3}}{2}\rmi & 0 \\ 0 & -\frac{\sqrt{3}}{2}\rmi & 0 & \rmi \\ \frac{\sqrt{3}}{2}\rmi & 0 & -\rmi & 0 \end{pmatrix},\\
 \label{mat7} \bar{J}_z &= S J_z S = \begin{pmatrix} \frac{3}{2} & 0&0 & 0\\ 0& -\frac{3}{2} &0 &0\\ 0& 0& -\frac{1}{2} &0 \\ 0& 0& 0& \frac{1}{2} \end{pmatrix}.
\end{align}
Defining the matrices $\bar{V}_i$ and $\bar{\gamma}$ as in the main text with $J_i\to \bar{J}_i$ we find
\begin{align}
  \label{mat8} \bar{V}_1 =\mathbb{1}_2\otimes \sigma_1,\ \bar{V}_2 = -\mathbb{1}_2\otimes \sigma_2,\ \bar{V}_3=\mathbb{1}_2\otimes \sigma_3,
\end{align}
or 
\begin{align}
\label{mat9} \bar{V}_i=\mathbb{1}\otimes \sigma_i^*,
\end{align}
This clearly shows that the representation of the Clifford algebra that specifies the Hamiltonian for $\alpha=0$ is of the ``first type''. Furthermore, the matrix $\bar{\gamma}_{45}$ that enters the time-reversal operator $\bar{\mathcal{T}}=\bar{\gamma}_{45}\mathcal{K}$ reads
\begin{align}
 \label{mat10} \bar{\gamma}_{45} = \mathbb{1}_2 \otimes \sigma_2,
\end{align}
whereas we have
\begin{align}
 \label{mat11} \bar{\mathcal{W}} = \bar{\gamma}_{12} = \sigma_2\otimes \mathbb{1}_2.
\end{align}

\section{Effective Weyl Hamiltonian and monopole charge}
We first construct the effective $2\times 2$ Weyl Hamiltonian at the nodes $\textbf{p}_n$, $n=1,\dots,4$, in the chiral topological semimetal phase with $\alpha>0$. The two orthogonal zero modes of $H_{\rm mf}(\textbf{p}_1)$ read
\begin{align}
 \label{weyl1} |0_1\rangle = \frac{1}{\sqrt{6}}\begin{pmatrix}\rmi\sqrt{3} \\ 1-\rmi \\ 1 \\ 0 \end{pmatrix},\ |0_1'\rangle=\mathcal{T}|0_1\rangle = \frac{1}{\sqrt{6}}\begin{pmatrix} 0 \\ \rmi \\ 1-\rmi \\ \sqrt{3} \end{pmatrix},
\end{align}
with similar expressions for $|0_n\rangle$ and $|0_n'\rangle$. From this we construct the projected Hamiltonian for momenta $\textbf{p}=\textbf{p}_n+\delta\textbf{p}$ close to one of the nodes according to
\begin{align}
\nonumber  H_0^{(n)} &= \begin{pmatrix} \langle 0_n|H_{\rm mf}(\textbf{p}_n+\delta \textbf{p})|0_n\rangle & \langle 0_n|H_{\rm mf}(\textbf{p}_n+\delta \textbf{p})|0_n'\rangle \\ \langle 0_n'|H_{\rm mf}(\textbf{p}_n+\delta \textbf{p})|0_n\rangle & \langle 0_n'|H_{\rm mf}(\textbf{p}_n+\delta \textbf{p})|0_n'\rangle  \end{pmatrix}\\
\label{weyl2}  &=  \begin{pmatrix} \langle 0_n|H(\delta\textbf{p})|0_n\rangle & \langle 0_n|H(\delta\textbf{p})|0_n'\rangle \\ \langle 0_n'|H(\delta\textbf{p})|0_n\rangle & \langle 0_n'|H(\delta\textbf{p})|0_n'\rangle  \end{pmatrix}.
\end{align}
Note that the $n$-dependence only results from the $n$-dependence of $|0_n\rangle$ and $|0'_n\rangle$ due to the linearity of the Hamiltonian. We arrive at $H_0^{(n)}=v^{(n)}_{ij}\delta p_i\sigma_j$ with
\begin{align}
 \nonumber v^{(1)} &=\frac{1+2\alpha}{3}\begin{pmatrix} 1 & 0 & 0 \\ 0 & 1 & 0 \\ 0 & 0 & 1 \end{pmatrix} + \frac{1-\alpha}{3} \begin{pmatrix} 0 & 1 & 1 \\ 1 & 0 & 1 \\ 1 & 1 & 0 \end{pmatrix},\\
 \nonumber v^{(2)} &=\frac{1+2\alpha}{3}\begin{pmatrix} 1 & 0 & 0 \\ 0 & 1 & 0 \\ 0 & 0 & 1 \end{pmatrix} + \frac{1-\alpha}{3} \begin{pmatrix} 0 & 1 & -1 \\ 1 & 0 & -1 \\ -1 & -1 & 0 \end{pmatrix},\\
 \nonumber v^{(3)} &=\frac{1+2\alpha}{3}\begin{pmatrix} -1 & 0 & 0 \\ 0 & -1 & 0 \\ 0 & 0 & 1 \end{pmatrix} + \frac{1-\alpha}{3} \begin{pmatrix} 0 & 1 & 1 \\ 1 & 0 & -1 \\ -1 & 1 & 0 \end{pmatrix},\\
 \label{weyl3} v^{(4)} &=\frac{1+2\alpha}{3}\begin{pmatrix} -1 & 0 & 0 \\ 0 & -1 & 0 \\ 0 & 0 & 1 \end{pmatrix} + \frac{1-\alpha}{3} \begin{pmatrix} 0 & 1 & -1 \\ 1 & 0 & 1 \\ 1 & -1 & 0 \end{pmatrix}.
\end{align}
We have 
\begin{align}
\mbox{det}(v^{(n)})=\alpha^2
\end{align}
for all $n=1,\dots,4$. The resulting monopole charge at the node $\textbf{p}_n$ is $q_n = \sgn[\mbox{det}(v^{(n)})]=+1$ for $\alpha>0$.

In the Weyl semimetal phase with $m\neq 0$ we consider $H_V(\textbf{p}_n)$ with $n=\text{a},\text{b}$. Since the x- and y-components of the nodal points vanish, we have a diagonal mean-field Hamiltonian at the nodes, namely
\begin{align}
 \label{weyl5} H_V(\textbf{p}_{\text{a}}) &= \frac{2m\alpha}{1+\alpha}\begin{pmatrix} 0 & 0 & 0 & 0 \\ 0 & -1 & 0 & 0  \\ 0 & 0 & 1 & 0 \\ 0 & 0 & 0 & 0 \end{pmatrix},\\ \label{weyl6} H_V(\textbf{p}_{\text{b}}) &= \frac{2m\alpha}{1-\alpha}\begin{pmatrix} -1 & 0 & 0 & 0 \\ 0 & 0 & 0 & 0 \\ 0 & 0 & 0 & 0 \\ 0 & 0 & 0 & 1 \end{pmatrix}.
\end{align}
The zero modes $|0_{n=\text{a},\text{b}}\rangle, |0'_{n=\text{a},\text{b}}\rangle=\mathcal{T}|0_{n=\text{a},\text{b}}\rangle$ immediately follow from this and we again define the projected Hamiltonian as in Eq. (\ref{weyl2}). We find $H_0^{(n)}=\tilde{v}^{(n)}_{ij}\delta p_i\sigma_j$ with
\begin{align}
\label{weyl7} \tilde{v}^{(\text{a})} &=\frac{1}{2}\begin{pmatrix} 0 & -(2-\alpha) & 0 \\ -(2-\alpha) & 0 & 0 \\ 0 & 0 & 2(1+\alpha)\end{pmatrix},\\
 \label{weyl8} \tilde{v}^{(\text{b})} &= \frac{1}{2}\begin{pmatrix} 0 & 2+\alpha & 0 \\ -(2+\alpha) & 0 & 0 \\ 0 & 0 & -2(1-\alpha)\end{pmatrix}.
\end{align}
We have
\begin{align}
 \label{weyl9} \mbox{det}(\tilde{v}^{(\text{a})}) &= -\frac{1}{4}(2-\alpha)^2(1+\alpha),\\
 \label{weyl10} \mbox{det}(\tilde{v}^{(\text{b})}) &= -\frac{1}{4}(2+\alpha)^2(1-\alpha),
\end{align}
so that the monopole charges are given by $q_{\text{a}}=\sgn[\mbox{det}(\tilde{v}^{(\text{a})})]=-1$ and $q_{\text{b}}=\sgn[\mbox{det}(\tilde{v}^{(\text{b})})]=\sgn(\alpha-1)$.

\section{Details of the renormalization group analysis}

\subsection{Perturbative propagator}
In order to determine the perturbative propagator $G_0(Q)$ we need to invert
\begin{align}
G_0^{-1}(Q) = A=   \rmi q_0\mathbb{1}_4  + q_i (V_i + \alpha U_i)
\end{align}
with frequency $q_0$. For arbitrary $\alpha$ this can be achieved with the help of the Cayley--Hamiltonian theorem which implies that the inverse of the $4\times 4$ matrix $G_0^{-1}$ is given by
\begin{align}
 \nonumber G_0(Q) ={}& \frac{1}{\mbox{det}(A)}\Biggl[\frac{1}{6}\Bigl([\mbox{tr} A]^3-3 (\mbox{tr}A)\mbox{tr}(A^2)+2\mbox{tr}(A^3)\Bigr)\mathbb{1}_4\\
 &-\frac{1}{2}\Bigl([\mbox{tr}A]^2-\mbox{tr}(A^2)\Bigr)A+(\mbox{tr}A)A^2-A^3\Biggr].
\end{align}
The electric charge enters the RG beta functions only through the combination 
\begin{align}
g_1'=g_1+\frac{e^2}{2}.
\end{align}
This is related to the fact that $G_0(Q)$ satisfies
\begin{align}
\label{intGG} \int_{q_0,\Omega} G_0(Q)^2 =0,
\end{align}
where $\int_{q_0}$ and $\int_{\Omega}$ denote the frequency and angular integration, respectively. Indeed, given Eq. (\ref{intGG}), the same reasoning as in Eqs. (A101)-(A104) of Ref \cite{PhysRevB.95.075149} can be applied to show this feature.

\subsection{Short-range interactions}\label{AppShort}

The RG flow of the couplings $\bar{g}_i$ is determined by the same procedure as laid out for Luttinger semimetals in Appendix 5 of Ref. \cite{PhysRevB.95.075149}. We confine our analysis to local point-like interaction terms. To incorporate the most general four-fermion interaction we write the interaction part of the Lagrangian as
\begin{align}
\label{model6b} L_{\rm short}=\sum_{A=1}^{16} \bar{g}_A (\psi^\dagger \Sigma^A\psi)^2,
\end{align}
where $\Sigma^A$ constitutes an $\mathbb{R}$-basis of Hermitean $4\times 4$ matrices satisfying $\mbox{tr}(\Sigma^A\Sigma^B)=4\delta^{AB}$. The symmetry properties of $H$ dictate which of the 16 entries of $\{\Sigma^A\}$ are independent under RG.

In the rotation invariant case (i.e. for $\alpha=2$) we have
\begin{align}
 \nonumber  L_{\rm short}^{(\rm rot)} ={}& \bar{g}_1(\psi^\dagger\psi)^2 + \bar{g}_2(\psi\gamma_a\psi)^2 \\
 \label{model7} &+ \bar{g}_\mathJ(\psi^\dagger \mathJ_i \psi)^2+\bar{g}_W(\psi^\dagger W_\mu \psi)^2,
\end{align}
where $\mathJ_i$, $\gamma_a$,\ $W_\mu$ are the three, five, and seven components of the irreducible $\text{SO}(3)$-invariant first-, second-, third-rank tensors made from products of the $J_i$. They read
\begin{align}
 \label{model8} \mathJ_i &= \frac{2}{\sqrt{5}}J_i,
\end{align}
and
\begin{align}
 \label{model9} \gamma_1 &= \frac{1}{\sqrt{3}}(J_x^2-J_y^2),\  \gamma_2 = J_z^2-\frac{5}{4}\mathbb{1}_4,\\
 \label{model11} \gamma_3 &= \frac{1}{\sqrt{3}}\{J_x,J_z\},\  \gamma_4 = \frac{1}{\sqrt{3}}\{J_y,J_z\},\\
 \label{model13}  \gamma_5 &=\frac{1}{\sqrt{3}}\{J_x,J_y\},
\end{align}
and
\begin{align}
 \label{model14} W_1 &= \frac{2\sqrt{5}}{3}\Bigl(J_{x}^3-\frac{41}{20}J_{x}\Bigr),\\
 \label{model15} W_2 &= \frac{2\sqrt{5}}{3}\Bigl(J_{y}^3-\frac{41}{20}J_{y}\Bigr),\\
 \label{model16} W_3 &= \frac{2\sqrt{5}}{3}\Bigl(J_{z}^3-\frac{41}{20}J_{z}\Bigr),\\
 \label{model17} W_4 &= \frac{1}{\sqrt{3}}\{J_x,(J_y^2-J_z^2)\},\\
 \label{model18}  W_5 &= \frac{1}{\sqrt{3}}\{J_y,(J_z^2-J_x^2)\},\\
 \label{model19} W_6 &= \frac{1}{\sqrt{3}} \{J_z,(J_x^2-J_y^2)\},\\
 \label{model20}  W_7 &= \frac{2}{\sqrt{3}}(J_xJ_yJ_z+J_zJ_yJ_x).
\end{align}
Note that 
\begin{align}
\mathcal{W}=W_7.
\end{align}
The matrices $\gamma_a$ are chosen such that $\gamma_{1,2,3}$ are real and $\gamma_{4,5}$ are imaginary. For a very detailed discussion of the decomposition of interaction vertices in Eq. (\ref{model7}), also in the cubic symmetric case, we refer to Ref. \cite{PhysRevB.95.075149}, where an identical notation was used to study systems described by a $4\times 4$ quadratic band touching Hamiltonian. Obviously the momentum dependence of $H$ does not affect the form of $L_{\rm short}$ and so all observations made in the mentioned reference are valid here as well.

For $\alpha\neq 2$ the system is invariant under the rotational cubic group $\text{O}$ only, and the irreducible tensors under $\text{SO}(3)$ need to be subdivided into irreducible representations of $O$. We write
\begin{align}
 \label{model21} \vec{E}=\begin{pmatrix} \gamma_1 \\ \gamma_2 \end{pmatrix},\ \vec{T}= \begin{pmatrix} \gamma_3 \\ \gamma_4 \\ \gamma_5 \end{pmatrix},\ \vec{W}=\begin{pmatrix} W_1 \\ W_2 \\ W_3 \end{pmatrix},\ \vec{W}' = \begin{pmatrix} W_4 \\ W_5 \\ W_6\end{pmatrix}.
\end{align}
The set $\{\mathbb{1},E_a,T_a,\mathJ_i,W_i,W'_i,W_7\}$ constitutes an appropriate basis of interactions in the cubic case. However, we are free to replace the elements $\mathJ_i$ and $W_i$ by
\begin{align}
 \label{model22} \vec{V} &= \frac{1}{\sqrt{5}}(\vec{\mathcal{J}}+2\vec{W}),\\
 \label{model23} \vec{U} &= \frac{1}{\sqrt{5}}(2\vec{\mathcal{J}}-\vec{W}),
\end{align}
which are the same matrices $\vec{V}$ and $\vec{U}$ as they appear in $H$. The set of matrices
\begin{align}
 \label{model24} \{\Sigma^A\} = \{ \mathbb{1},\ E_a,\ T_a,\ V_i,\ U_i,\ W'_i, W_7\}
\end{align}
then comprises a computationally advantageous orthogonal $\mathbb{R}$-basis of Hermitean $4\times 4$ matrices with $\mbox{tr}(\Sigma^A\Sigma^B) = 4\delta_{Ab}$. The seven elements in Eq. (\ref{model24}) allow us to construct seven distinct insulating ordering channels, namely
\begin{align}
 L_1 &=\bar{g}_1 (\psi^\dagger \psi)^2,\\
 L_2 &=\bar{g}_2 (\psi^\dagger\vec{E}\psi)^2,\\
 L_3 &=\bar{g}_3 (\psi^\dagger\vec{T}\psi)^2,\\
 L_4&=\bar{g}_4 (\psi^\dagger\vec{V}\psi)^2,\\
 L_5 &=\bar{g}_5 (\psi^\dagger\vec{U}\psi)^2,\\
 L_6 &=\bar{g}_6 (\psi^\dagger\vec{W}^\prime\psi)^2,\\
 L_7 &=\bar{g}_7 (\psi^\dagger W_7\psi)^2.
\end{align}
Only three of these expressions are linearly independent, and we choose to parametrize the interaction in $L_{\rm int}$ by $L_{1,2,3}$. The remaining four are related to these by the Fierz identities
\begin{align}
 L_4 &= -\frac{3}{2}L_1-\frac{3}{2}L_2+\frac{1}{2}L_3,\\
 L_5 &= -\frac{3}{2}L_1 -\frac{1}{2}L_3,\\
 L_6 &= -\frac{3}{2}L_1-\frac{1}{2}L_3,\\
 L_7 &= -\frac{1}{2}L_1 +\frac{1}{2}L_2-\frac{1}{2}L_3.
\end{align}
Note that $L_5=L_6$. In the rotationally invariant case we have $\bar{g}_2=\bar{g}_3$. Each individual term $L_1,\dots,L_7$ is invariant under $\psi \to \mathcal{W}\psi$ since we have the (anti)commutation relations
\begin{align}
 [\mathcal{W},V_i] &= [\mathcal{W},T_i] = 0,\\
 \{\mathcal{W},U_i\} &=\{\mathcal{W},E_a\} = \{\mathcal{W},W'_i\}=0.
\end{align} 
Consequently, the invariance of $L_{\rm short}$ under this transformation is independent of the choice of Fierz basis.

We may alternatively express $L_{\rm short}$ in terms of the superconducting channels of the system. The number of such terms is identical to the number of Fierz-inequivalent insulating channels. We have
\begin{align}
 L_{\rm short} = \bar{g}_{\rm s} L_{\rm s} + \bar{g}_{\rm d,E}L_{\rm d,E} + \bar{g}_{\rm d,T} L_{\rm d,T}
\end{align}
with
\begin{align}
 L_{\rm s} ={}& (\psi^{\dagger} \gamma_{45}\psi^*)(\psi^{\rm T}\gamma_{45}\psi)\\
 L_{\rm d,E} &= \sum_{a=1}^2(\psi^{\dagger} \gamma_a\gamma_{45}\psi^*)(\psi^{\rm T}\gamma_{45}\gamma_a\psi)\\
  L_{\rm d,T} &=\sum_{a=3}^5(\psi^{\dagger} \gamma_a\gamma_{45}\psi^*)(\psi^{\rm T}\gamma_{45}\gamma_a\psi),
\end{align} 
with the linear relation \cite{PhysRevB.93.205138,PhysRevB.95.075149}
\begin{align*}
 L_{\rm s} &= \frac{1}{4}(L_1+2L_2+3L_3),\\
 L_{\rm d,E} &= \frac{1}{4}(L_1-3L_3),\\
 L_{\rm d,T} &=\frac{1}{4}(L_1-2L_2-L_3).
\end{align*}
In the rotationally symmetric case ($\alpha=2$) we have $g_{\rm d,E}=g_{\rm d,T}$, again reducing the number of independent couplings to two.

\subsection{Susceptibility exponents and admixture $\kappa(\alpha)$}\label{AppSusc}

The susceptibility exponents determine which order parameter condenses at the quantum critical point described by a certain RG fixed point. We define the susceptibility exponent $\eta=\eta_M$ of a fermion bilinear $\psi^\dagger M \psi$ or $\psi^\dagger M \psi^*$ through the scaling dimension
\begin{align}
\label{susc1} [h_M] = z +\eta_M,
\end{align}
where $h=h_M$ is introduced by coupling a term $L_M = h_M (\psi^\dagger M \psi^{(*)})$ to the Lagrangian. For a detailed discussion which fully applies here see App. 6 of Ref. \cite{PhysRevB.95.075149}. In our case $z=1$ is the trivial dynamic critical exponent due to $e=0$ at the fixed point. Importantly, for each fixed point we have to test every ordering channel individually and determine the one with the largest susceptibility. On the other hand, due to cubic symmetry, we can restrict to the ten distinct cubic channels. For this purpose we couple
\begin{align}
  L_{\mathbb{1}} &= h_{\mathbb{1}} (\psi^\dagger \psi),\\
 L_{E} &= h_{E} (\psi^\dagger \gamma_1\psi),\\
  L_{T} &= h_{T} (\psi^\dagger \gamma_3\psi),\\
  L_{V} &= h_{V} (\psi^\dagger V_1\psi),\\
  L_{U} &= h_{U} (\psi^\dagger U_1\psi),\\
 L_{W'} &= h_{W'} (\psi^\dagger W_4\psi),\\
  L_{\mathcal{W}} &= h_{\mathcal{W}} (\psi^\dagger W_7\psi),\\
 L_{\rm s} &= h_{\rm s} (\psi^\dagger \gamma_{45}\psi^*),\\
  L_{\rm d,E} &= h_{\rm d,E} (\psi^\dagger \gamma_1 \gamma_{45}\psi^*),\\ 
L_{\rm d,T} &= h_{\rm d,T} (\psi^\dagger \gamma_3\gamma_{45}\psi^*)
\end{align}
to the Lagrangian and determine the corresponding flow equations $\dot{h}_M=(z+\eta_M)h_M$ to read off the susceptibilities. 

To give some example we present the susceptibilities for the analytically tractable cases $\alpha=0$ and $\alpha=2$. For $\alpha=0$ we have
\begin{align}
\nonumber \eta_{\mathbb{1}}&=\eta_{E} =\eta_{\mathcal{W}}=0,\ \eta_T = \frac{2}{3}(g_1'-2g_2-5g_3),\\
\nonumber  \eta_V& =\frac{2}{3}(g_1'+2g_2-g_3),\ \eta_U= \eta_{W'}=\frac{2}{3}(g_1'+g_3),\\
\nonumber \eta_{\rm s} &= -(g_1'+2g_2+3g_3),\ \eta_{\rm d,E} =-(g_1'-3g_3),\\
 \label{susc2} \eta_{\rm d,T} &= \frac{1}{3}(-g_1'+2g_2+g_3).
\end{align}
For the rotation invariant case with $\alpha=2$ we have
\begin{align}
 \nonumber \eta_{\mathbb{1}} &=0,\ \eta_E = \frac{1}{5}(g_1'-4g_2-3g_3),\\ 
 \nonumber \eta_T &= \frac{1}{5}(g_1'-2g_2-5g_3),\ \eta_{W'} = \frac{43}{105}(g_1'+g_3)\\
 \nonumber \eta_{\mathcal{W}} &= \frac{43}{105}(g_1'-2g_2+3g_3),\ \eta_{\rm s} = -\frac{2}{3}(g_1'+2g_2+3g_3),\\
 \label{susc3}\eta_{\rm d,E} &=-\frac{1}{3}g_1'+g_3,\ \eta_{\rm d,T} = -\frac{1}{3}(g_1'-2g_2-g_3).
\end{align}
We omitted $\eta_V$ and $\eta_U$ here for a reason that will be explained in the next paragraph. Further imposing rotation invariance onto the couplings by setting $g_2=g_3$ we obtain
\begin{align}
 \nonumber \eta_{\mathbb{1}} &=0,\ \eta_{\gamma_a} = \eta_E = \eta_T = \frac{1}{5}(g_1'-7g_2),\\
 \nonumber \eta_{\mathcal{J}} &= \frac{4}{15}(g_1'+g_2),\ \eta_W =\eta_{W'}=\eta_{\mathcal{W}} = \frac{43}{105}(g_1'+g_2),\\
 \label{susc4} \eta_{\rm s} &= -\frac{2}{3}(g_1'+5g_2),\ \eta_{\rm d}=\eta_{\rm d,E}=\eta_{\rm d,T} = -\frac{1}{3}g_1'+g_2.
\end{align}
As expected, the susceptibilities of different components of same-rank tensors in Eq. (\ref{model6b}) coincide in the rotation invariant limit.

In most cases, by coupling a term $h_M(\psi^\dagger M \psi)$ to the Lagrangian while setting $h_N=0$ for all other matrices $N\neq M$, we only generate a running of the coupling $h_M$. However, if there is a cubic transformation that relates $M$ and $N$, this is no longer true. In our case, coupling a term $h_V (\psi^\dagger V_i\psi)$ to the Lagrangian generates a term $h_U (\psi^\dagger U_i\psi)$, and vice versa. Referring once more to Ref. \cite{PhysRevB.95.075149}, Eqs. (A110)-(A114) for details, we briefly review here how to determine the correct scaling behavior.

We write
\begin{align}
 \nonumber h(\psi^\dagger V_i\psi) \Rightarrow  h(\psi^\dagger V_i\psi)+h\Bigl[\eta_V (\psi^\dagger V_i\psi) +a (\psi^\dagger U_i\psi)\Bigr],\\
 h(\psi^\dagger _iU\psi) \Rightarrow h(\psi^\dagger U_i\psi)+ h\Bigl[b (\psi^\dagger V_i\psi) +\eta_U (\psi^\dagger U_i\psi)\Bigr],
\end{align}
where ``$\Rightarrow$'' stands for ``generates a term under RG''. In general, $a$ and $b$ depend on $\alpha$ and the couplings $g_i$. We find that for all values of $\alpha$ we have
\begin{align}
 a &= -K(g_1'+g_3),\\
 b&= -K(g_1'+2g_2-g_3)
\end{align}
with $K=K(\alpha)>0$ a positive constant. Some values are:
\begin{center}
\begin{tabular}{|c|c|c|c|c|c|c|c|c|}
\hline $\alpha$ & 0 & 0.5 & 1 & 1.5 & 2 & 2.296 & 2.5 & 5 \\ 
\hline $K$ & 0 & 0.12 & 0.25 & 0.106 & 0.057 & 0.0425 & 0.035 & 0.0077  \\ 
\hline 
\end{tabular} 
\end{center}
We observe that $K$ is small and vanishes for $\alpha\to0$ and $\alpha\to \infty$, with a maximum around $\alpha \approx 1$.

We introduce $M_i = c V_i + c' U_i$ with real coefficients such that $c^2+c'{}^2=1$. Obviously, only the ratio $\kappa=c'/c$ is of relevance. The maximal (and also the minimal) susceptibility $\eta_M$ will then come from a linear combination that satisfies the self-consistent relation
\begin{align}
 \label{susc5} h(\psi^\dagger M_i\psi) \Rightarrow h(1+\eta_{M})(\psi^\dagger M_i \psi)
\end{align}
with $\eta_{M} =\eta_V + \frac{c'}{c}b$ and
\begin{align}
 0 &\stackrel{!}{=} c'(\eta_U-\eta_V) +\frac{1}{c}\Bigl(c^2 a - c'{}^2 b\Bigr).
\end{align}
Te last equation is solved by
\begin{align}
 \kappa &= \frac{c'}{c} = \frac{\eta_U-\eta_V}{2b}\pm\sqrt{\frac{(\eta_U-\eta_V)^2}{4b^2}+\frac{a}{b}},\\
 \eta_{M} &=\frac{\eta_V+\eta_U}{2} \pm \frac{1}{2}\sqrt{(\eta_U-\eta_V)^2+4ab}.
\end{align}
Since we are after the largest susceptibilities, we are interested in the ``$+$'' solution of $\eta_{M}$. This corresponds to choosing ``$-$'' in $\kappa$ at fixed point V, and ``$+$'' in $\kappa$ at fixed point W. For V we have $|\kappa| < 0.05$ with the largest value around $\alpha\sim 1.25$. Consequently, up to a tiny correction we have
\begin{align*}
 M_i \approx V_i
\end{align*}
at fixed point V. In contrast, at W we find that $\kappa\gg1$ (of order 10) for all $\alpha$, and so we can regard this case as
\begin{align*}
 M_i \approx U_i.
\end{align*} 
The exponent $\eta_{M}$ at W is larger than $\eta_{\mathcal{W}}=3$ for $\alpha \leq 0.7$.

\begin{figure}[t!]
\centering
\begin{minipage}{0.48\textwidth}
\includegraphics[width=8cm]{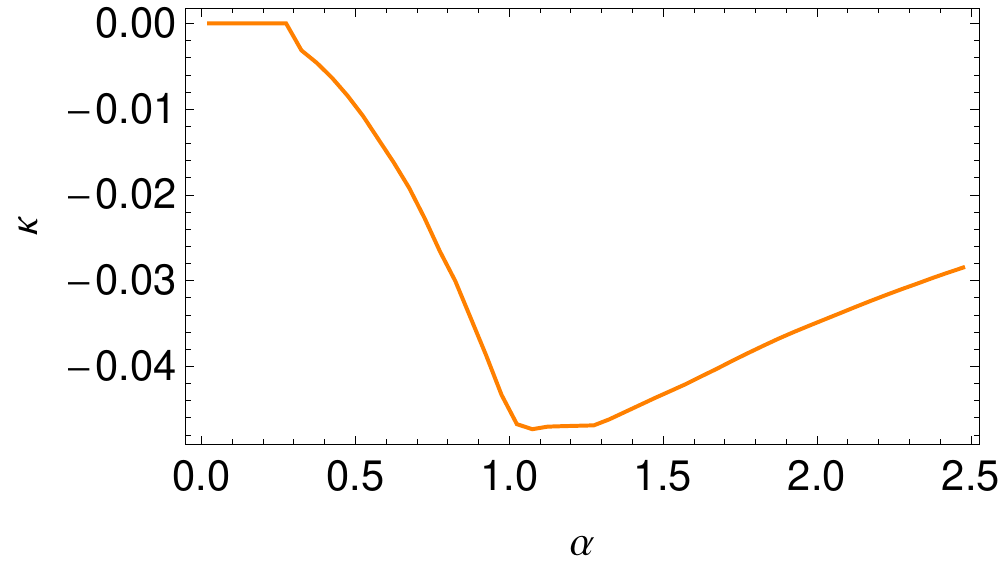}
\includegraphics[width=8cm]{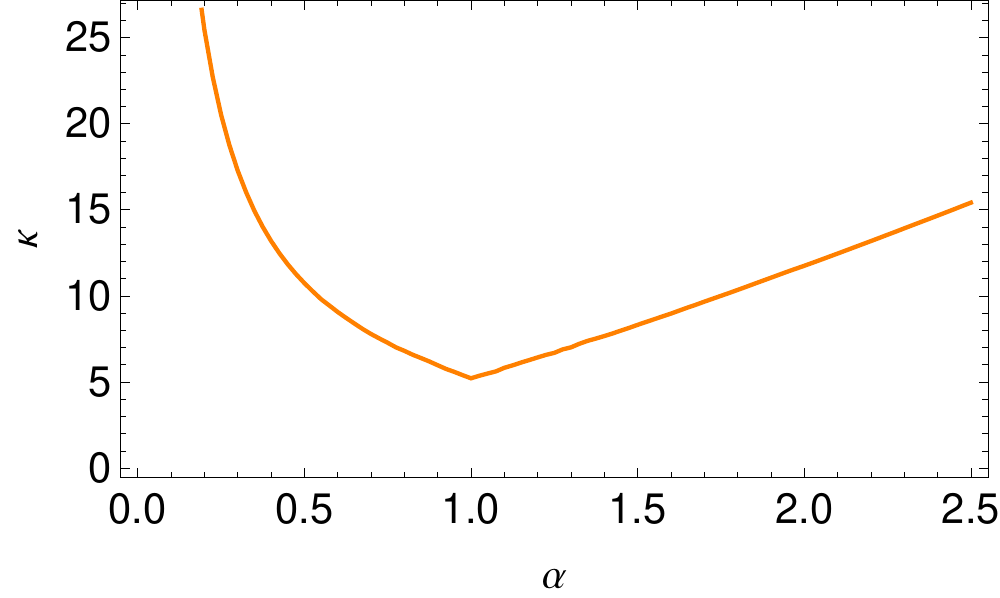}
\caption{The ratio $\kappa=c'/c$ that maximizes $\eta_M$ for the bilinear $\psi^\dagger(cV_i+c'U_i)\psi$ at the fixed point V (upper panel) and W (lower panel). \emph{Upper panel.} At V we have $|\kappa|<5\%$ and so the order parameter is to a very good approximation given by $\langle \psi^\dagger V_i\psi\rangle$. \emph{Lower panel.} At W the ratio $\kappa\gg1$ is large and so the corresponding order parameter is approximately $\langle \psi^\dagger U_i \psi\rangle$. However, for $\alpha\geq 0.7$, the leading instability at W is the condensation of $\chi=\langle \psi^\dagger \mathcal{W}\psi\rangle$ since $\eta_{\mathcal{W}}=3$ is the larger susceptibility.}
\label{Figkappa}
\end{minipage}
\end{figure}

\subsection{Flow equations at high-symmetry points}

\noindent The flow equations for $\alpha=0$ read
\begin{align}
 \label{rg3} \dot{g}_1 &= -2g_1 -g_2^2-6g_2g_3-5g_3^2,\\
 \label{rg4} \dot{g}_2 &=-2g_2+g_2^2-2g_2g_3-3g_3^2,\\
 \label{rg5} \dot{g}_3 &= -2g_3 -\frac{5}{3}g_2^2-\frac{14}{3}g_2g_3-3g_3^2.
\end{align}
Note that, apart from the trivial term $-2g_1$ in the first line, $g_1$ and $e$ are absent in these equations. We find the quantum critical points SC and V given by
\begin{align}
 \label{rg6} \text{SC}:\ (g_1,g_2,g_3)_\star &= \Bigl(-\frac{7}{16},-\frac{3}{16},-\frac{5}{16}\Bigr),\\
 \label{rg7} \text{V}:\ (g_1,g_2,g_3)_\star &= \Bigl(\frac{1}{2},\frac{3}{2},-\frac{1}{2}\Bigr).
\end{align} 
The largest susceptibility exponents at SC and V are $\eta_{\rm s}=7/4=1.75$ and  $\eta_V=8/3=2.67$, respectively.

In the rotation invariant case ($\alpha=2$) we have
\begin{align}
\nonumber  \dot{g}_1 ={}&-2g_1 -\frac{2}{15}g_1'g_2 -\frac{1}{5}g_1'g_3 -\frac{76}{105}g_2^2 \\
 \label{rg8} &- \frac{164}{35} g_2g_3 -\frac{79}{35}g_3^2,\\
 \nonumber \dot{g}_2={}& -2g_2 +\frac{12}{35}g_1'g_2-\frac{19}{35}g_1'g_3-\frac{4}{15}g_2^2\\
 \label{rg9} &-\frac{58}{35}g_2g_3-\frac{15}{7}g_3^2,\\
 \nonumber \dot{g}_3 ={}& -2g_3 -\frac{38}{105}g_1'g_2+\frac{17}{105}g_1'g_3 -\frac{24}{35}g_2^2\\
 \label{rg10} &-\frac{272}{105}g_2g_3-\frac{83}{105}g_3^2.
\end{align}
The quantum critical points are given by 
\begin{align}
 \label{rg11} \text{SC}:\ (g_1,\ g_2,\ g_3)_\star &= (-0.841,\ -0.450,\ -0.450),\\
 \label{rg12} \text{W}:\ (g_1,\ g_2,\ g_3)_\star &= (1.22,\ -1.22,\ 1.22),\\
 \label{rg13} \text{V}:\ (g_1,\ g_2,\ g_3)_\star &= (3.31,\ 3.07,\ -1.20).
\end{align}
the leading susceptibility exponents are given by $\eta_{\rm s}=2.062$,\ $\eta_{\mathcal{W}}=3$ and $\eta_{V_i}= 4.06$, respectively. Note that the fixed point W satisfies $g_1=-g_2=g_3$ and $\eta_{\mathcal{W}}=d$. For $g_2=g_3$ the flow equations read
\begin{align}
 \label{rg14} \dot{g}_1 &= -2g_1 -\frac{1}{3}g_1'g_2-\frac{23}{3}g_2^2,\\
 \label{rg15} \dot{g}_2 &= -2g_2 -\frac{1}{5}g_1'g_2-\frac{61}{15}g_2^2.
\end{align}
This set of equations only supports the superconducting quantum critical point SC. We should think of the fixed points W and V in this limit as approached in an anisotropic system with $\alpha\to 2$ from above or from below.

\section{Cubic and tetrahedral symmetry group}\label{AppCubic}
In this appendix we summarize some properties of the rotational cubic point group which are relevant for our analysis. We begin by deriving explicit expressions for the group elements and then discuss the rotational or chiral tetrahedral point group.

The cubic point group $O_h$ consists of the transformations that leave a three-dimensional cube invariant. Clearly, inversion is such a symmetry transformation. All other elements of the group can be expressed as a rotation or a rotation combined with an inversion. The ``rotational'' subgroup $O$, which is the symmetry group of our problem at hand, consists of the true rotations. Whereas $O_h$ consists of 48 elements, there are 24 elements in $O$. 

In the following we consider the three-dimensional representation of the group acting on position or momentum space vectors $\textbf{x}=(x_1,x_2,x_3)^T$ or $\textbf{p}=(p_1,p_2,p_3)^T$, respectively. The matrices $J_i$ also transform like a vector under $O$. The group $O$ is a subgroup of $\text{SO}(3)$, and so every element $R\in O$ satisfies $R^TR=\mathbb{1}_3$ and $\mbox{det}(R)=1$. Every rotation can be specified by an axis $\textbf{n}=(n_1,n_2,n_3)^T$ and a rotation angle $\vphi$ according to
\begin{align}
 \label{cub1} R(\textbf{n},\vphi) = \mathbb{1}_3 + \sin\vphi K + (1-\cos \vphi) K^2
\end{align}
with
\begin{align}
\label{cub2} K = \begin{pmatrix} 0 & -n_z & n_y \\ n_z & 0 & -n_x \\ -n_y & n_x & 0 \end{pmatrix}.
\end{align}
The inverse of $R(\textbf{n},\vphi)$ is $R(-\textbf{n},\vphi)$, and the self-inverse elements of $O$ are precisely the ones represented by symmetric matrices, which corresponds to rotations by $0^{o}$ or $180^{o}$. If $R$ is an element of $O$, then $IR$ with inversion $I=\text{diag}(-1,-1,-1)$ is the corresponding element of $O_h$ that includes an inversion.

\begin{widetext}
We now summarize the 24 group elements of $O$. The unit element is $R_1=\mathbb{1}_3$. We further have the following symmetry operations:\\
\noindent{\emph{(i) Three rotations by $180^o$ about a 4-fold axis.}} The corresponding rotation axes are the $x,y,z$-axes connecting opposite faces given by $\textbf{n}=(1,0,0)^T, (0,1,0)^T, (0,0,1)^T$, which leads to the self-inverse group elements
\begin{align}
 \label{cub3} R_2 = \begin{pmatrix} 1 & 0 & 0 \\ 0 & -1 & 0 \\ 0 & 0 & -1 \end{pmatrix},\ R_3 = \begin{pmatrix} -1 & 0 & 0 \\ 0 & 1 & 0 \\ 0 & 0 & -1 \end{pmatrix},\ R_4 = \begin{pmatrix} -1 & 0 & 0 \\ 0 & -1 & 0 \\ 0 & 0 & 1 \end{pmatrix}.
\end{align}
\emph{(ii) Eight rotations by $120^o$ about a 3-fold axis.} These 3-fold axes are the axes connecting opposite vertices. The sign of $\textbf{n}$ matters and we have
\begin{align} 
 \label{cub4} \textbf{n}=\frac{1}{\sqrt{3}}\begin{pmatrix} 1 \\ 1 \\ 1 \end{pmatrix},\ \frac{1}{\sqrt{3}}\begin{pmatrix} 1 \\ -1 \\ 1 \end{pmatrix},\ \frac{1}{\sqrt{3}}\begin{pmatrix} -1 \\ 1 \\ 1 \end{pmatrix},\ \frac{1}{\sqrt{3}}\begin{pmatrix} -1 \\ -1 \\ 1 \end{pmatrix},
\end{align} 
which leads to
\begin{align}
 \label{cub5} R_5 = \begin{pmatrix} 0 & 0 & 1 \\ 1 & 0 & 0 \\ 0 & 1 & 0 \end{pmatrix},\  R_{6} = \begin{pmatrix} 0 & -1 & 0 \\ 0 & 0 & -1 \\ 1 & 0 & 0 \end{pmatrix},\ R_{7} = \begin{pmatrix} 0 & -1 & 0 \\ 0 & 0 & 1 \\ -1 & 0 & 0 \end{pmatrix},\ R_{8}=\begin{pmatrix} 0 & 0 & -1 \\ 1 & 0 & 0 \\ 0 & -1 & 0 \end{pmatrix}.
\end{align}
The inverse elements follow from $\textbf{n}\to -\textbf{n}$ and read
\begin{align}
 \label{cub6} R_{9}=R_{5}^T,\ R_{10}=R_{6}^T,\ R_{11}=R_{7}^T,\ R_{12}=R_{8}^T.
\end{align}
\emph{(iii) Six rotations by $90^o$ about a 4-fold axis.} These 4-fold axes are again the $x,y,z$-axes connecting opposite faces, but this time the sign of $\textbf{n}$ matters. We have $\textbf{n}=(1,0,0)^T, (0,1,0)^T, (0,0,1)^T$, which leads to
\begin{align}
 \label{cub7} R_{13} &= \begin{pmatrix} 1 & 0 & 0 \\ 0 & 0 & -1 \\ 0 & 1 & 0 \end{pmatrix},\  R_{14} = \begin{pmatrix} 0 & 0 & 1 \\ 0 & 1 & 0 \\ -1 & 0 & 0 \end{pmatrix},\ R_{15}= \begin{pmatrix} 0 & -1 & 0 \\ 1 & 0 & 0 \\ 0 & 0 & 1 \end{pmatrix},
\end{align}
and their inverses with $\textbf{n}\to -\textbf{n}$ and
\begin{align}
 \label{cub8} R_{16}=R_{13}^T,\ R_{17}=R_{14}^T,\ R_{18}=R_{15}^T.
\end{align}
\emph{(iv) Six rotations by $180^o$ about a 2-fold axis.} The axes are the axes connecting opposite edges.  The matrices are their own inverses and the sign of $\textbf{n}$ does not matter. We have
\begin{align}
 \label{cub9} \textbf{n} = \frac{1}{\sqrt{2}}\begin{pmatrix} 1 \\ 1 \\ 0 \end{pmatrix},\ \frac{1}{\sqrt{2}}\begin{pmatrix} 1 \\ -1 \\ 0 \end{pmatrix},\ \frac{1}{\sqrt{2}}\begin{pmatrix} 1 \\ 0 \\ 1 \end{pmatrix},\ \frac{1}{\sqrt{2}}\begin{pmatrix} 1 \\ 0 \\ -1 \end{pmatrix},\ \frac{1}{\sqrt{2}}\begin{pmatrix} 0 \\ 1 \\ 1 \end{pmatrix},\ \frac{1}{\sqrt{2}}\begin{pmatrix} 0 \\ 1 \\ -1 \end{pmatrix},
\end{align}
which leads to
\begin{align}
 \label{cub10} R_{19} &= \begin{pmatrix} 0 & 1 & 0 \\ 1 & 0 & 0 \\ 0 & 0 & -1 \end{pmatrix},\ R_{20} = \begin{pmatrix} 0 & -1 & 0 \\ -1 & 0 & 0 \\ 0 & 0 & -1 \end{pmatrix},\ R_{21} = \begin{pmatrix} 0 & 0 & 1 \\ 0 & -1 & 0 \\ 1 & 0 & 0 \end{pmatrix},\\
 \label{cub11} R_{22} &= \begin{pmatrix} 0 & 0 & -1 \\ 0 & -1 & 0 \\ -1 & 0 & 0 \end{pmatrix},\ R_{23} = \begin{pmatrix} -1 & 0 & 0 \\ 0 & 0 & 1 \\ 0 & 1 & 0 \end{pmatrix},\ R_{24} = \begin{pmatrix} -1 & 0 & 0 \\ 0 & 0 & -1 \\ 0 & -1 & 0 \end{pmatrix}.
\end{align}
\end{widetext}

The largest subgroup of $O$ is the rotational tetrahedral group $T$ with 12 elements. It consists of the rotations that leave a tetrahedron invariant. (Note that the tetrahedral group $T_d$ is commonly defined such that it includes inversion as well, and so it has 24 elements.) The elements of $T$ are precisely the first twelve $R_{1,\dots,12}$ in our notation. The group $T$ leaves the expressions $J_1J_2J_3+J_3J_2J_1$ and $p_1p_2p_3$ invariant. In contrast, the remaining twelve elements of $O$ change the sign of these expressions.

\end{document}